\documentclass[amsmath,amssymb,amsfonts,aps,pre,showpacs,floatfix,twocolumn]{revtex4}
\usepackage{graphicx}
\usepackage{amsmath,amssymb,amsfonts}
\usepackage{natbib}
\usepackage{times}
\usepackage{color}
\newcommand{\NOTER}[1]{#1}
\newcommand{\NOTEC}[1]{#1}
\newcommand{\NOTEK}[1]{#1}
\newcommand{\mum}{{\mu}\text{m}}
\newcommand{\del}{\nabla}

\begin{document}
\pacs{47.57.J-, 47.61.Jd, 05.40.Jc, 82.70.Dd}
\title{Nonlinear Elasticity of Microsphere Heaps}

\author{Carlos P. Ortiz, Karen E. Daniels, Robert Riehn} 
\affiliation{Dept. of Physics, North Carolina State University, Raleigh, NC, USA}

\date{\today}

\begin{abstract}
Thermal fluctuations, geometric exclusion, and external driving all govern the mechanical response of dense particulate suspensions. Here, we measure the stress-strain response of quasi-two-dimensional flow-stabilized microsphere heaps in a regime in which all three effects are present using a microfluidic device. We observe that the elastic modulus \NOTEC{and the mean interparticle separation} of the heaps \NOTEC{are} tunable via the confining stress provided by the fluid flow. Furthermore, the measured stress-strain curves exhibit a universal nonlinear shape which can be predicted from a thermal van der Waals equation of state with excluded volume. \NOTEC{This analysis indicates that many-body interactions contribute a significant fraction of the stress supported by the heap}.  

\end{abstract}

\maketitle

\section{Introduction}
 
The emergent mechanical properties of particulate matter play a central role in
many physical processes and biological systems.
Hard-sphere Brownian suspensions allow direct study of important condensed matter phenomena, such as the emergence of irreversibility from reversible equations of motion~\cite{Corte2008,Pine2005}, the glass transition~\cite{Weeks2000,Yethiraj2003, Ballesta2008,Hunter2012}, crystal nucleation and dislocation~\cite{Nelson1979, Schall2006}, and dynamical heterogeneities~\cite{Abate2007,Nordstrom2011}. Both thermal fluctuations and geometric exclusion play significant roles in determining the bulk response of particulate matter, but little is known about the interplay between the effects. To probe this regime, we have developed a novel in-situ technique for measuring the elastic response of quasi-two-dimensional microsphere heaps confined by fluid stresses within a microfluidic device.

While the rheological properties of dense colloidal suspensions have been the
subject of much  study \cite{Krieger1972, VanderWerff1989, DeKruif1985,
VanderWerff1999, Chen2010,Sentjabrskaja2013,Eisenmann2010, Mattsson2009, 
Larsen2010,Michailidou2009,Mutch2013,Pham2002,Segre1995}, 
and experiments show that elastic behavior dominates over viscous behavior on approach to a 
critical packing density~\cite{Mason1995,Sentjabrskaja2013}, much less is known
about the solid-like properties of such systems above a rigidity transition. 
In this paper, we quantify the elastic response of very soft colloidal
solids by varying the fluid stress on microsphere heaps assembled within a 
microfluidic channel~\cite{Ortiz2013}.  This system is
particularly suited to measuring the mechanical response of colloidal
assemblies: the size and shape of the microsphere heap are direct observables,
the relative contribution of fluid forces and thermal diffusion is tunable via
the choice of particle size and fluid flow rate,
and confining stresses from the fluid are applied as a bulk effect rather than
from a single boundary.
Inertial effects are negligible (thin film Reynolds number
Re~$\approx 10^{-8}$), and thermal and athermal effects are approximately equal
in magnitude (P\'eclet number Pe~$\gtrsim 1$).
Here, we show that microsphere solids are more elastically-robust than 
ordinary solids, and their elastic modulus 
increases with the confining fluid stress. We observe that this
family of solids is characterized by a universal nonlinear stress-strain curve,
and we interpret this finding in light of a thermal 
equation of state~\citep{vanderWaals1873}.

\begin{figure*}
\includegraphics[width=\linewidth]{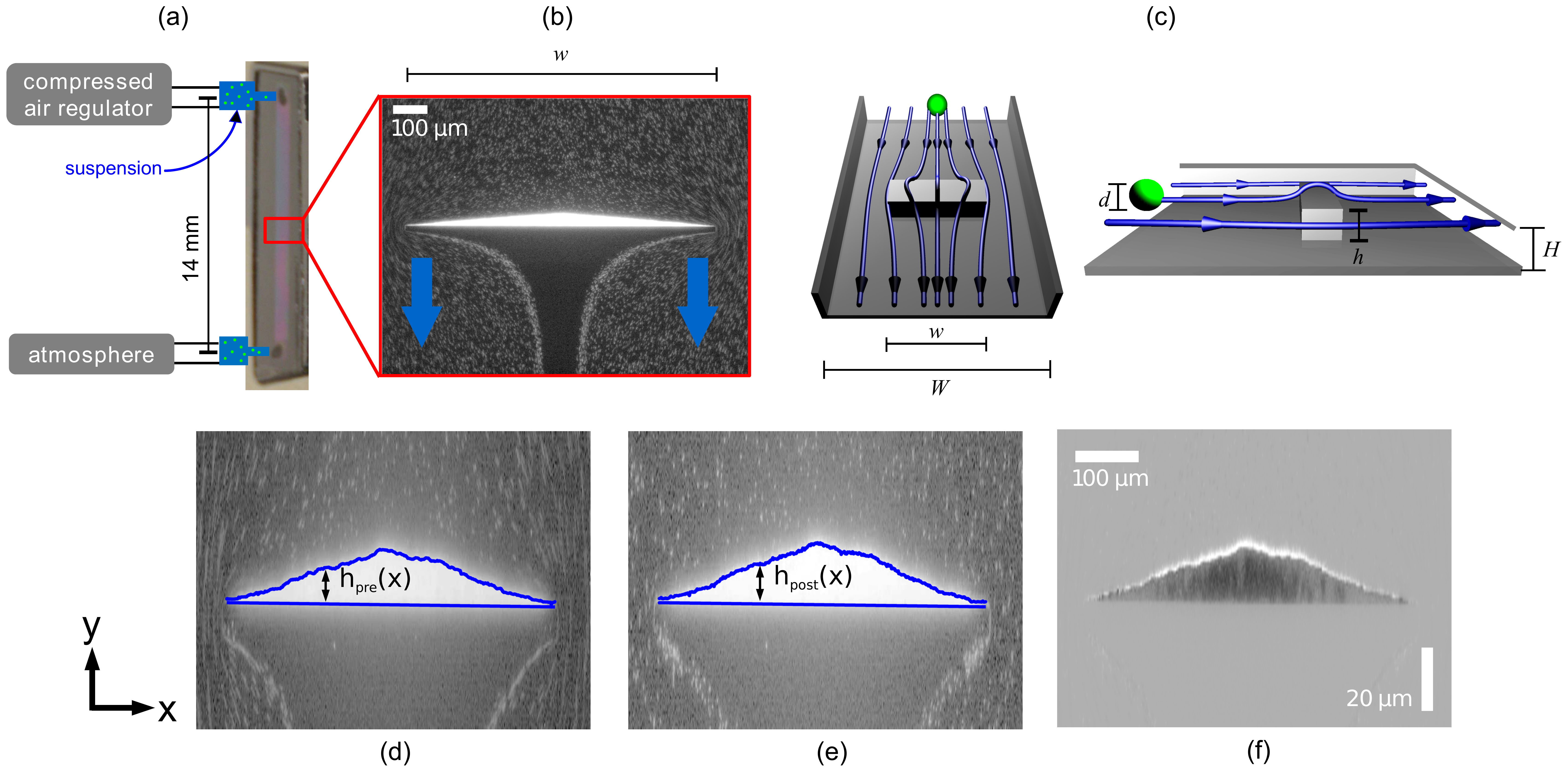}
\caption{ Overview of Deformation Experiment. 
(a) Microfluidic experimental setup.
(b) Grayscale fluorescence microscopy image depicting steady-state microsphere heap with suspension flowing around it. 
(c) Top-view and side-view diagram of Hele-Shaw microfluidic device geometry  with
fluid flow shown by blue streamlines.
$W \times H \times L = 
1\,\mathrm{mm} \times 0.94\,\mu\mathrm{m}  \times 14\,\mathrm{mm}$, not to scale), showing a
single microsphere flowing towards the flat-topped barrier
($w \times h \times \ell = 
563\,\mu$m$\times0.84\,\mu$m$\times 10\, \mu$m.
(a) Schematic microchip geometry 
Steady-state raw fluorescence images (d) pre-perturbation and (e) 2.0 s
after a perturbation of amplitude $\Delta P = -3.5$~kPa. Linear
grayscale represents fluorescence intensity, and blue line is the detected
heap boundary. Fluid flow direction is from top to bottom.
(f) Image-difference of the same sequence. }
\label{fig:heaps}
\end{figure*}

\section{Experiment}

\subsection{Apparatus}

We use a microfluidic device to assemble microsphere heaps by flowing a dilute, 
aqueous suspension against a barrier (Fig.~\ref{fig:heaps}). The particles form a
quasi-two-dimensional heap through gradual accumulation against a
barrier, with the steady-state angle and size of the flow-stabilized
solid set by the fluid flow rate~\cite{Ortiz2013}. 

Particles are deposited from a dilute suspension with mean two-dimensional areal density $\rho$, 
given by the number of microspheres per unit area, of ($110/(100~{\mu}\text{m})^2$)
against a flat-topped barrier.
Due to the weak polydispersity of the particles, we observe 
polycrystalline domains within our heaps when they are examined with a higher-magnification
objective.
The device channel has a quasi-2D geometry with height $H = 1.8~d$,
accommodating  only a single layer of particles. 
The barrier is $h = 1.5~d$ high and occupies the center $w = 1000~d$ of a channel
of width $W = 2000\,d$, thus permitting flow around the top and sides of the
barrier.

The aqueous suspension is composed of fluorescent polystyrene 
microspheres (530~nm, 5.9\% polydispersity,
elastic modulus $\approx 4$~GPa), measured using dynamic light scattering (Malvern zetasizer) to have diameter $d=530$~nm.
We use steric and electrostatic stabilization (sulfate functionalized surface
with $\zeta$-potential of $-60$~mV and coated with Triton X-100) to provide
reversible inter-particle and channel-particle interactions.  
The surfactant concentration was chosen to be small enough to avoid creating micelles, and this was confirmed using dynamic light scattering. Therefore,  interparticle interactions beyond 2 nm are dominated by the electrostatic (DLVO) interaction.

For the experiments reported here,  we prepare heaps under varying hydrodynamic 
confinement  and record video
of the heap size/shape due to changes in the applied pressure, from
which we calculate the elastic bulk modulus $K$.  
We have prepared heaps under a variety of different barrier geometries,  pH of the medium, and particle/channel ratios, and find that heap elasticity is a general phenomenon. In the experiments described below, we focus on a single set of parameters.

We control the suspension flow rate through the channel with a digital pressure regulator (AirCom PRE1-UA1) that applies pressurized air ($P_0 = 0-10$~kPa above
atmosphere) to an o-ring sealed reservoir at the device inlet; the outlet is maintained at atmospheric pressure. We assemble heaps at steady pressures $P_0$ applied at the microchannel inlet such that $P_0 = 0-10$~kPa above atmospheric pressure. 
This pressure range includes the transition to stable heaps \cite{Ortiz2013}, and thus is the regime with  the largest possible influence of thermal fluctuations.
The corresponding  P\'eclet numbers are between 4 and 28, which corresponds
to observed heap angles of 1.3$^\circ$ to 5.4$^\circ$.
We deform the heap via a step change $\Delta P$ which, mediated by the flow rate, 
changes the fluid stress by an amount
$\Delta\sigma \propto \Delta P$. 
We use a $P$ to $\sigma$ calibration curve developed using a Darcy flow model (see Appendix \ref{sec:hyd}). 
Using fluorescence microscopy, we visualize the deformation of the heap in 
response to the imposed $\Delta \sigma$ in real time. We chose wide-field fluorescence microscopy rather than confocal detection as this allows recording  the fast dynamics of the entire system in real time.
The light source is a continuous-wave Xenon lamp (X-cite 120); images of
the fluorescence are collected with a Nikon Eclipse 80i fluorescence microscope
with a $10\times$ Plan Fluor objective (N.A. 0.8) and an Andor Luca emCCD camera operating
at 10 Hz frame rate.

\begin{figure}
\centering
\includegraphics[width=0.6\columnwidth]{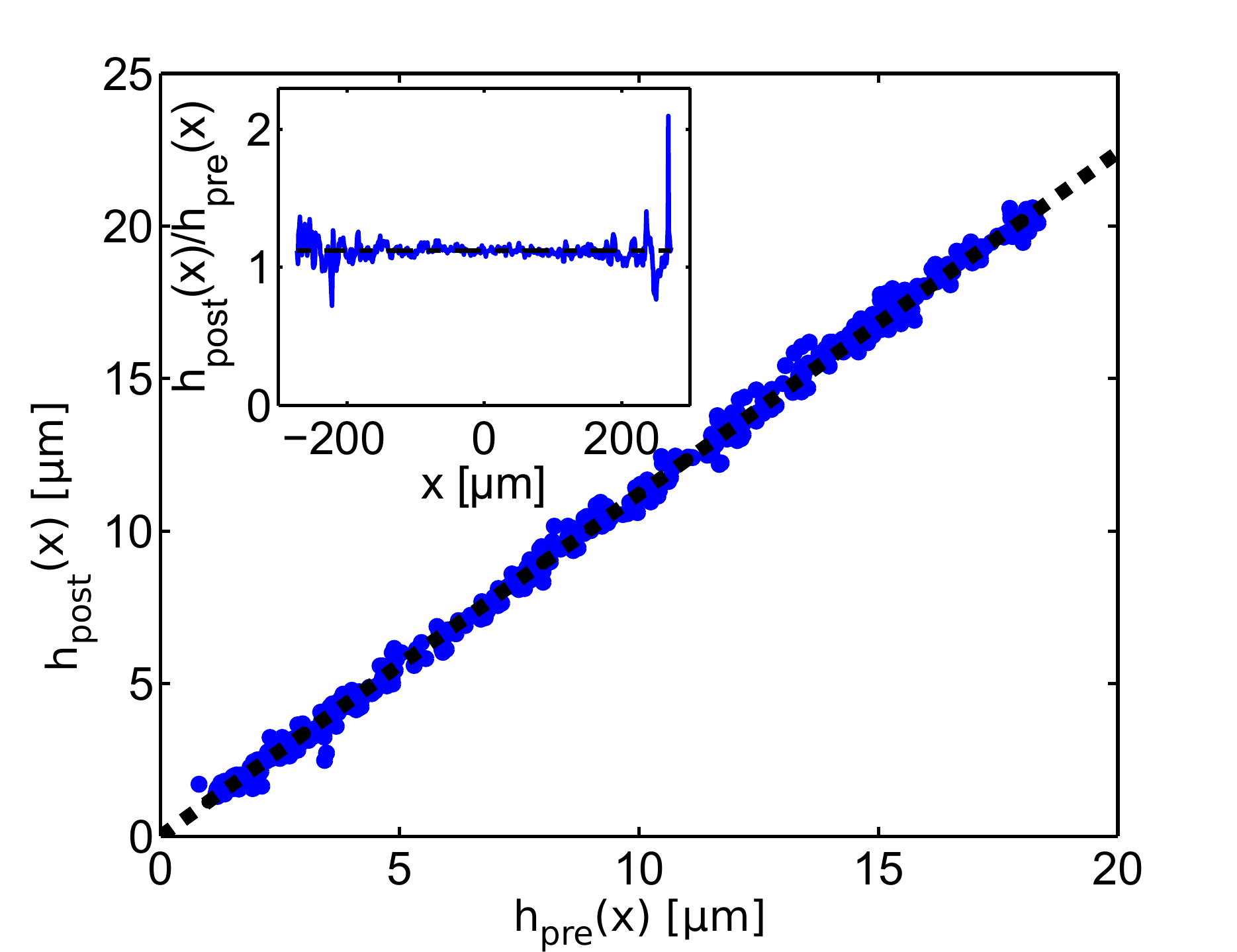}
\caption{
 Observation of uniform strain.
(a) Parametric plot of $h_\text{post}(x)$ versus $h_\text{pre}(x)$ with
inset local strain $\gamma(x)=h_\text{post}(x)/h_\text{pre}(x)$. In both plots,
the dashed line corresponds to a mean strain of 12\%, measured from
the slope of (a).}
\label{fig:localstrain}
\end{figure}

To quantify the strain, we locate the
boundary profile $h(x)$ of the heap by computing the gradient of the image along
$\hat y$, then using a Gaussian fit to extract the heap profile. 
The pre- and post-deformation profiles
$h_\text{pre}(x)$ and $h_\text{post}$, provide a means to
measure the local strain. As shown in Fig.~\ref{fig:localstrain}, a parametric
plot of these two profiles is observed to be linear; this indicates that the
deformation is well-described by a single mean strain.  Equivalently, the ratio
$h_\text{post}(x) / h_\text{pre}(x)$, shown in the inset, demonstrates that the
strain is constant along the $\hat{x}$-direction. Therefore, we can use the change in area, $\Delta A$, as a surrogate measure.

Since we are primarily interested in the elastic response, for which
the timescale is $\approx 1$~s, we are able to measure $\Delta A$ by taking the
difference between two linear regressions to $A(t)=\int\,h(x,t)\,dx$, fit during the 3~s
immediately before and after the deformation. This technique filters out any
slow drifts, and additionally allows measurements even when the deformation
amplitudes are comparable to the background fluctuations. We tested the
dependence of $\Delta A / A_0$ on its measurement early or late within the
time-series and found no significant effect.

Figs.~\ref{fig:heaps}bc show sample pre- and post-deformation images; the
difference image (Fig.~\ref{fig:heaps}d) shows a bright band of newly occupied
volume that indicates an expansion of the heap.
We locate the boundary profile of the heap via a Gaussian fit of the gradient in the $y$-direction. 
We find that applying a positive stress change $+\Delta\sigma$ leads to a compression of the heap in the $y$-direction. The resulting deformation is well-described by a
single mean strain given by $\frac{\Delta A}{A_0}$, the ratio of the change in the size of the heap, $\Delta A$, to the size of the heap prior to the deformation, $A_0$ (as shown in Fig.~\ref{fig:localstrain}).

\subsection{Inter-particle Separations \label{sec:parsep}}

\begin{figure}
\centerline{\includegraphics[width=0.8\columnwidth]{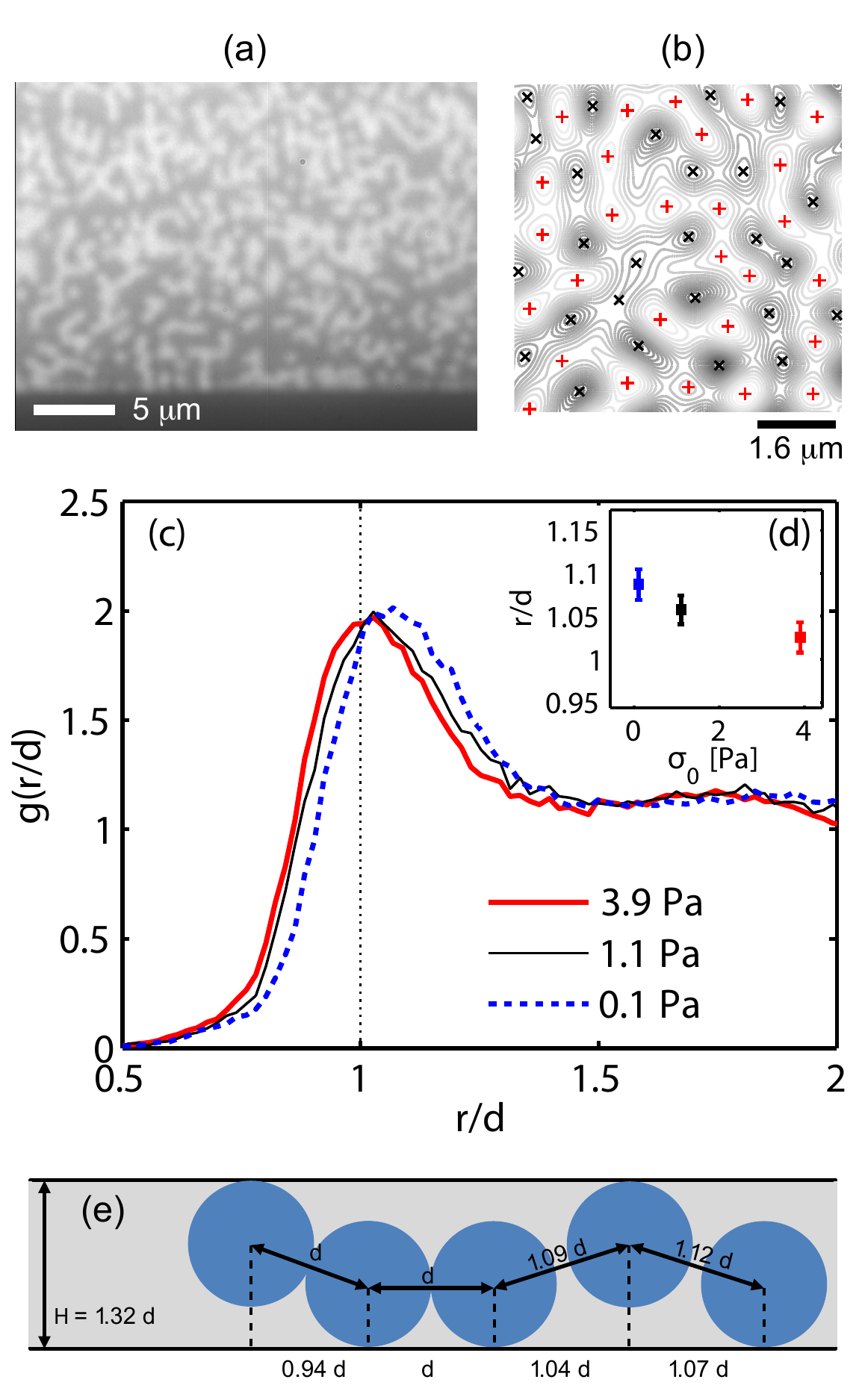}}
\caption{(a) Fluorescence microscopy image of a microsphere heap composed of bidisperse particles under steady confining stress $\sigma_0=0.1$~Pa. Diameter of dim particles is 600~nm and of bright particles is 710~nm.  (b) Intensity contour lines showing particle centroids in microsphere heap, detected using Wiener deconvolution. Light contours with $+$ (red) correspond to bright regions, and dark contours with $\times$ (black) correspond to dim regions.  
(c) Pair radial distribution functions, $g(r/d)$, measured at three different $\sigma_0$: 0.1 Pa, 1.1 Pa, 3.9 Pa. 
(d) Peak position as a function of the confining stress $\sigma_0$, measured from a parabolic fit to the peak, with 1-$\sigma$ (68.3 \%) confidence intervals. 
(e) Schematic of possible configurations of microspheres in the channel and their apparent separations when observed from above.
}
\label{fig:g}
\end{figure}

Due to the small size of the particles (530~nm) used in the experiment, it is not possible to optically resolve individual particles. Therefore, we perform a supplemental set of experiments in using a bidisperse 1:1 mixture of larger 600~nm and 710~nm particles ($H$=1.32~$d_g$) for which we could obtain particle positions. We use the larger diameter of 710~nm as the unit $d_g$ in the following analysis and discussion. This steady-state pack is under $\sigma_0=0.1$, 1.1, or 3.9 Pa (corresponds to $P_0 = 0.1$, 1, or 7 kPa.)
The pack is imaged with a $60\times$ water immersion objective (N.A. 1.00) and a $4\times$ beam expander. We compute a mean image by taking the average over a small set of images, each with a 0.1~s exposure time, for a 1.0~s integration time.  We use a Wiener deconvolution algorithm with a Gaussian point spread function to process the raw fluorescence images and find the particle positions. Particles are detected in the deconvolved image by searching for local maxima that are connected bright pixels of a threshold value relative to their environment. For 
each local maxima region, we find a centroid that corresponds to the most probable particle position. These centroids are shown as black crosses 
overlayed over the raw image in Fig.~\ref{fig:g}a. \NOTEC{We determine the accuracy of the algorithm on synthetic images with known particle centers and radii, subject to simulated Poisson noise and optical blurring. Using Wiener deconvolution, we are able to recover the position of the particles to $0.05~d_g$. We detect no pixel-biasing in either the results from synthetic or the real images. } From the centroids, we compute $g(r)$ from the normalized histogram of the interparticle pairwise distances. \NOTEC{Each $g(r)$ curve in Fig.~\ref{fig:g} contains more than 500,000 particle positions. }

From a parabolic fit for separations from $0.9~d_g$ to $1.2~d_g$, we find that the position of the nearest-neighbor peak in $g(r)$ is $(1.09 \pm 0.02)~d_g$ for a confining stress of 0.1~Pa. Increasing the confining stress reduces the interparticle separation, as shown in Fig.~\ref{fig:g}c. \NOTEK{Using the simulated images,} \NOTEC{we find errors in the particle positions do not systematically bias the peak position in $g(r)$ relative to the true position; the absolute standard error in the peak position is $0.02~d_g$, while the uncertainty is $0.03~d_g$.} \NOTEK{The trend in Fig.~\ref{fig:g}cd is larger than these errors.} In addition, we have verified that these results are robust to small changes in the image processing parameters and pair distribution algorithm parameters. The significance of this result is that most of the nearest neighbors are found 9\% farther than 1 particle diameter, and are thus not true contacts in the Hertzian sense. \NOTER{Our finding is strengthened by the 
observation that $g(r)$ is confining-stress dependent, which would not be anticipated for a Hertzian contact network.  \NOTEK{Finally,} $d_g$ is larger than the average particle diameter, and so the first peak in $g(r)$ would actually be larger if it were in units of the average particle diameter.}

Our observation of non-zero $g(r)$ below $1.0~d_g$ suggests a fraction of particle configurations with either intimate surface contact or staggered out-of-plane positions.  Fig.~\ref{fig:g}e shows schematic of the influence on $g(r)$ of microsphere configurations in a microchannel of $H=1.32~d_g$. Apparent separations below $1.0~d_g$ are possible due to staggered configurations with intimate surface contacts and that apparent separations of $1.07~d_g$ could indicate interparticle separations as large as $1.12~d_g$.

\begin{figure}
\centering{
\includegraphics[width=0.8\columnwidth]{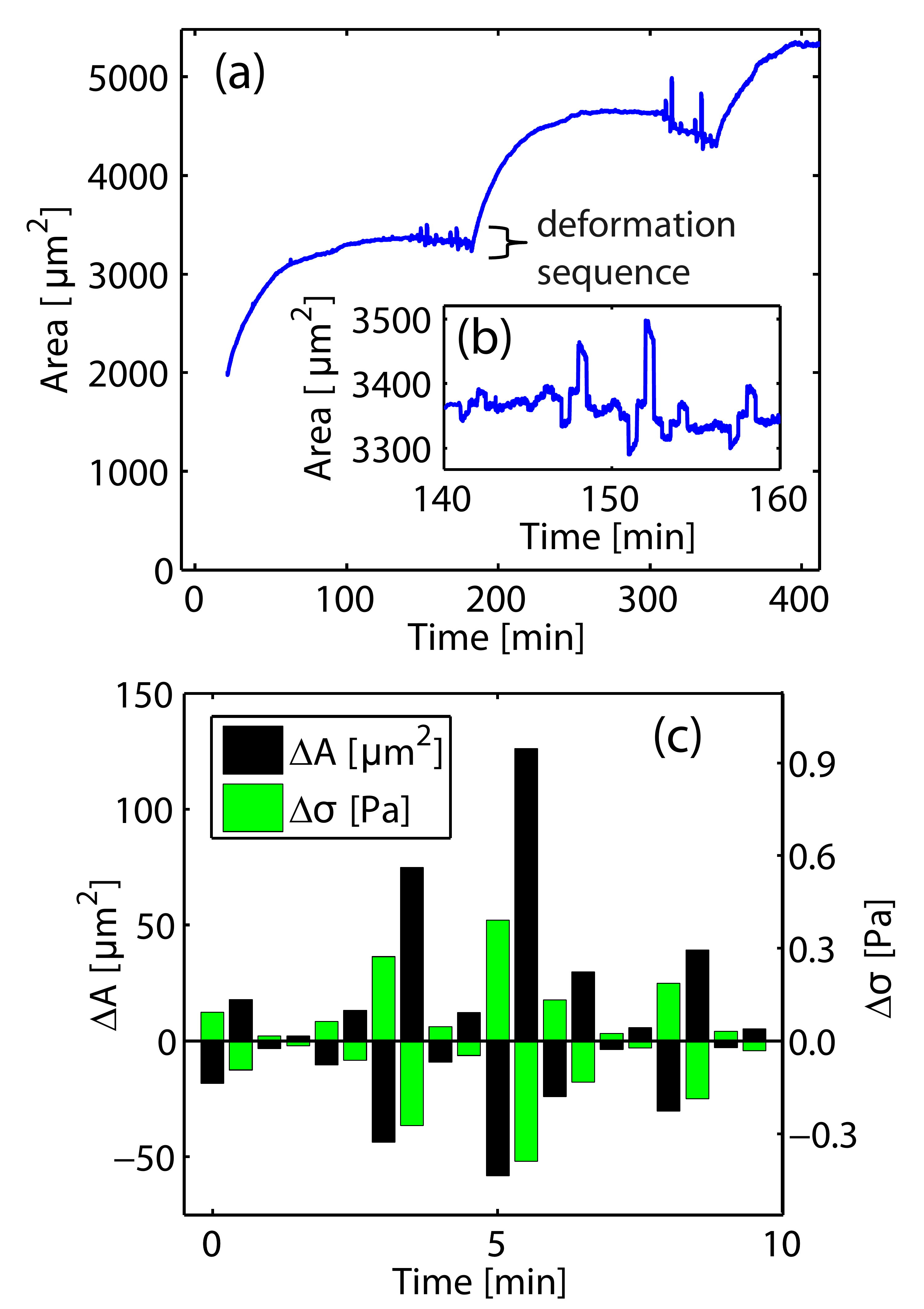}}
\caption{Stress-Strain Measurement Protocol. (a) Heap growth to steady state at $P_0 = 1$~kPa
($\sigma_0=0.7$~Pa)  during two hours of particle accumulation, followed by a
sequence of 80 randomized deformations, and then repeated for a larger
value of $P_0 = 2$~kPa. (b) Inset: Zoom of $A(t)$, for
which the extracted values of $\Delta A$ (black) and $\Delta \sigma$ (green) are
shown in (c). }
\label{fig:heapstimeseries}
\end{figure}
\section{Results}

We measure the elastic response for heaps formed at seven different initial
stresses ($\sigma_0 = 0.1$ to 4 Pa). Fig.~\ref{fig:heapstimeseries}a illustrates
the time-evolution of $A$ for the first two values of $\sigma_0$.  
At time $t=0$, the suspension is released into the microchannel and 
over 2 hours the
heap size grows to its steady state at $A_0$. This value is
$\sigma_0$-dependent, and was previously attributed to a critical P\'eclet
number~\cite{Ortiz2013}. A series of deformation experiments is then performed
on the steady-state heap; this process is repeated for increasing values of
$\sigma_0$, each of which is allowed to reach a steady state.

The series of deformation cycles on each heap is an identical randomized set of
80 logarithmically-spaced values of $\Delta \sigma$  from 2~mPa to 3.8~Pa.  
For each $\Delta \sigma$, the cycle contains four deformations around $\sigma_0$
($+\Delta \sigma, -\Delta \sigma , -\Delta \sigma, +\Delta \sigma$); overall the 
full series comprises 320 deformations.  Fig.~\ref{fig:heapstimeseries}b shows 
a detailed view of this sequence, and Fig.~\ref{fig:heapstimeseries}c
shows the measured values of $(\Delta \sigma, \Delta A)$ extracted from the sequence. 
As is expected for an elastic response, an
increase (decrease) in $\sigma$ corresponds to an decrease (increase) in $A$.  
Each pair of such measurements  provides a point along the stress-strain curve
at that value of $\sigma_0$.

We find that large-scale expansions lead to some shedding of particles, which
leads to lower $A$ after the deformation. The erosion process is facilitated by
the fluidization of particles at the heap interface~\cite{Ortiz2013}.  Because
re-deposition of particles is slower than the timescale of the experiments, we
observe an apparent slight shrinking of
the heap during a deformation sequence, as shown in the second deformation
series in Fig.~\ref{fig:heapstimeseries}a.  
While randomizing the sequence minimizes the impact of memory effects $\Delta A$
vs. $\Delta \sigma$ relationship, the erosion complicates the measurements for
small $\Delta\sigma$ late in the sequence. To account for this effect, we define
the effective heap size $A_0$ for each step immediately before the deformation
step rather than the initial steady state area.  

\begin{figure}
\includegraphics[width=\columnwidth]{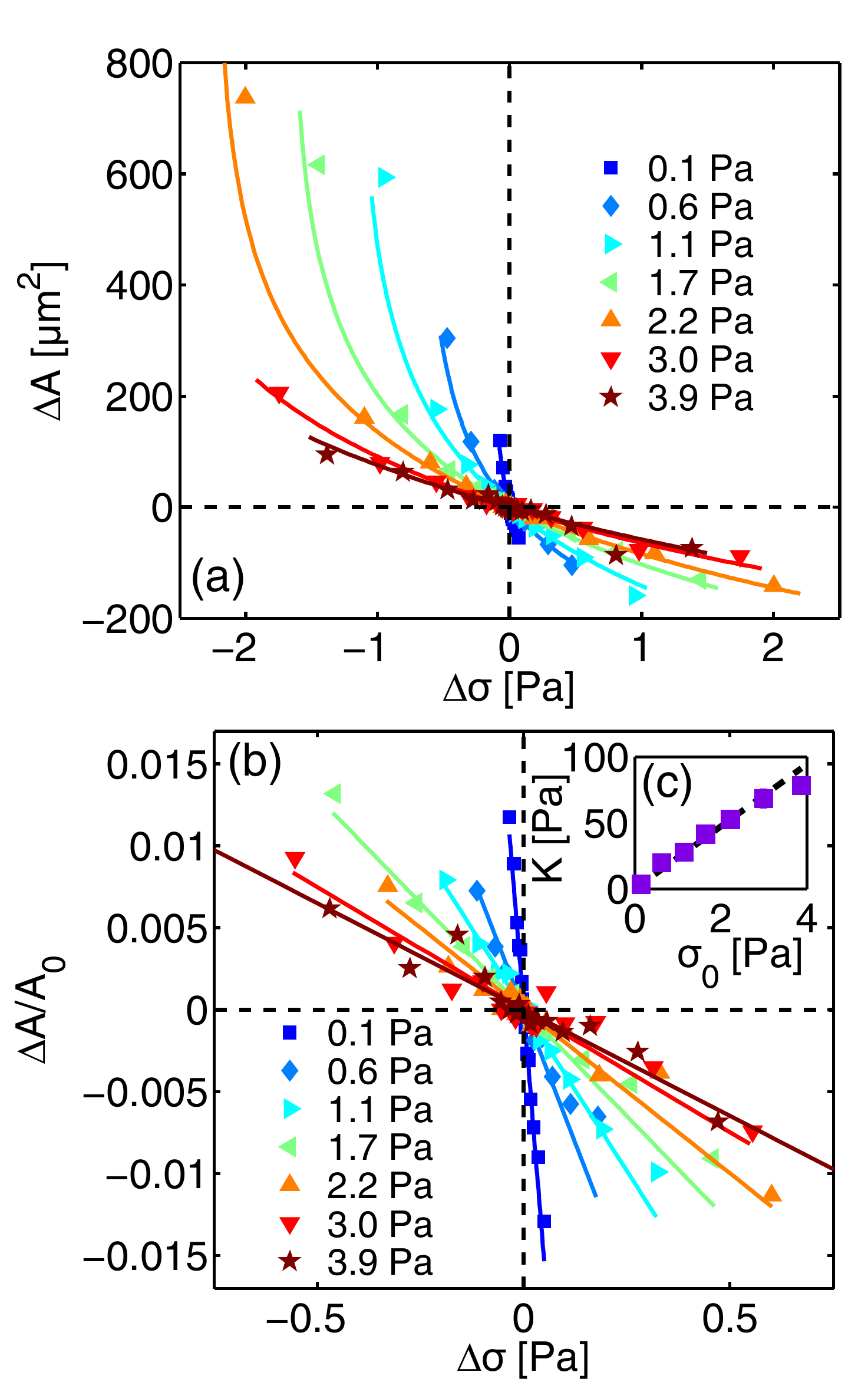}
\caption{ Raw stress-strain measurements. 
(a) Scatter plot of perturbation amplitude $\Delta \sigma$
vs. deformation amplitude $\Delta A$ for seven difference values of $\sigma_0$
indicated by symbol shape. Lines are fits to Eq.~\ref{eqn:gamma}.
The measurement range is largest at intermediate values of $\sigma_0$,
truncated for small values by the lower limit of heap formation at 
$\mathrm{Pe} \approx 2$~\cite{Ortiz2013} and at large values by the pressure
regulator.
(b) Stress-strain measurements for the same data. Each line is a
linear fit to determine the elastic modulus $K$ (see Eq.~\ref{eqn:Kdeffinite}).
(c) Inset: Measured values of $K$ from (b) as a function of  $\sigma_0$, with error
bars indicating one standard deviation  estimates obtained from the fits. The
dashed line is the linear fit is to Eq.~\ref{eqn:Kscale} with
$\alpha = 23.3 \pm 0.74$.
}
\label{fig:modulus}
\end{figure}

\subsection{Deformations}

We present the stress-strain curves for the full set of experiments in
Fig.~\ref{fig:modulus}a. Near $\Delta \sigma = 0$, we find an approximately 
linear response  for all steady-state heaps, as shown in
Fig.~\ref{fig:modulus}b. At large deformations, a nonlinear relationship
becomes apparent, but is nonetheless still reversible. For each curve,
we measure the linear elastic modulus $K$ by
linearizing around $\Delta \sigma = 0$:
\begin{equation}
\frac{\Delta A}{A_0} = - \frac{ \Delta \sigma}{K}
\label{eqn:Kdeffinite}
\end{equation}
over the range $|\Delta A / A_0| < 0.015$.  The magnitude of $K$ determined here
is consistent with the literature for crystallized sub-micron colloidal
polystyrene~\cite{Crandall1977}
or silica~\cite{Shinohara2001, Okubo2002, Murai2012} particles. The observation of a large
elastic regime (up to 10\%-15\% strain), stands in contrast to ordinary solids
in which such strains would cause fracture.

\begin{figure}
\includegraphics[width=\columnwidth]{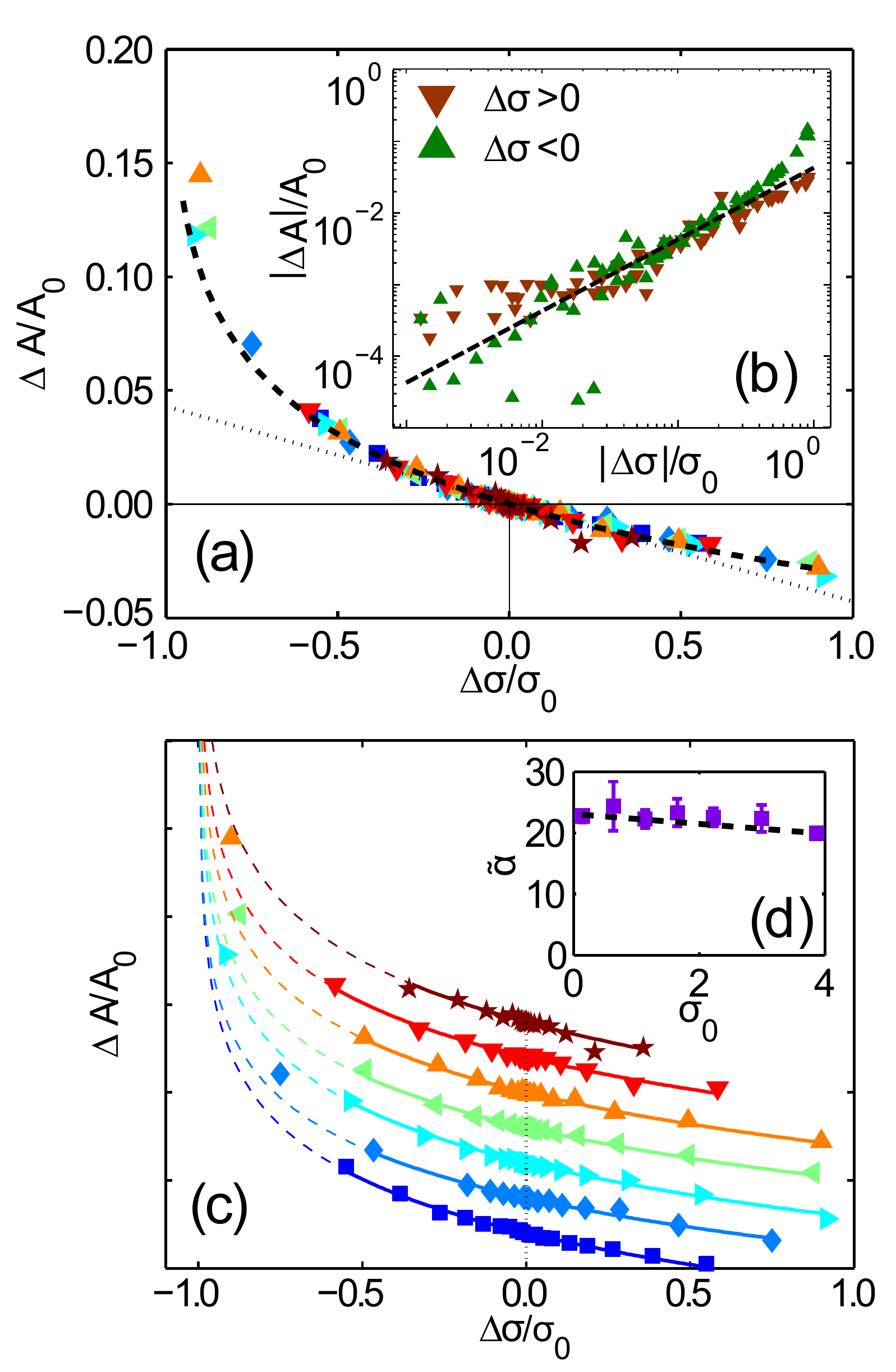}
\caption{Scaled stress-strain measurements. (a) Strain $\Delta A/A_0$ as a function of the scaled perturbation amplitude
$\Delta \sigma/\sigma_0$, showing data collapse. Dashed solid line is from
Eq.~\ref{eqn:gamma} with $\tilde{\alpha} = 22.5$; thin dotted line is
Eq.~\ref{eqn:Kdeffinite} with $K = \alpha\sigma_0$ and $\alpha = 23.3$. Symbol legend is
the same as in Fig.~\ref{fig:modulus}. (b) Inset: Log-log plot of $\Delta
A/A_0$ versus $\vert\Delta \sigma\vert/\sigma_0$
where compressions are upward (green) triangles and decompressions are downward
(brown) triangles.
Dashed solid line is Eq.~\ref{eqn:Kdeffinite} with $K = \alpha\sigma_0$ and
$\alpha=23.3$.
(c) The same data, with the  curves vertically shifted by 0.1 to allow for
comparison. Each curve is fit to Eq.~\ref{eqn:gamma} to find a best-fit value of 
$\tilde{\alpha}$; 
solid lines denote the fit region and the dashed lines are continuations beyond that region. 
(d) Inset: Values of
$\tilde{\alpha}$ for each dashed curve fit in (c). Error bars
are one standard deviation parameter variance estimates for each regression
obtained
from adding the parameter variance from the Hessian of the error function and
systematic errors. Dashed line is a linear fit:
$\tilde{\alpha}=\left(23.4\pm2.3\right)+\left(-0.24\pm0.17\right)\sigma_0$.}
\label{fig:lindensity}
\end{figure}

As shown in Fig.~\ref{fig:modulus}c, we find that flow-stabilized
solids formed under higher compression (larger $\sigma_0$) 
are stiffer (larger $K$). This is unlike
ordinary solids, where $K$ is a constant for small compressions, and suggests
that the elastic response is controlled by the steady state stress $\sigma_0$.
We observe an approximately linear relationship between $K$ and $\sigma_0$,
with 
\begin{equation}
K = \alpha \, \sigma_0.
\label{eqn:Kscale}
\end{equation}
where  $\alpha$ is the compressibility factor.
Because Eqns.~\ref{eqn:Kdeffinite} and \ref{eqn:Kscale} together imply $\frac{\Delta
A}{A_0} \propto \frac{\Delta \sigma}{\sigma_0}$ for small deformations, we seek to 
collapse all of the stress-strain measurements onto a
single curve. In Fig.~\ref{fig:lindensity}a, we test whether this dependence
extends to larger deformations by rescaling $\Delta \sigma$ by $\sigma_0$. In
Fig.~\ref{fig:lindensity}b, we observe that the linear approximation extends for two orders of
magnitude in strain, with increasing scatter below $\Delta A/A_0 \approx
0.02$ and deviations in decompression above $\Delta A/A_0 \approx
0.3$. The extent of this data collapse, well beyond the linear regime which
motivated it, suggests a deeper connection between the stress and
strain which we explore below.

\section{Discussion}


\NOTEK{In order to explain the observed universal, nonlinear elastic response, it is reasonable to consider either thermal (classical, hard-sphere equation of state~\cite{Planck1909,Reiss1959}) or athermal (jamming~\cite{Liu2010,VanHecke2010}) theories. In these experiments, we observe that because the particles are both  immersed in a fluid and subject to thermal fluctuations, they exhibit diffusive motion and are not typically in contact with each other. As shown in Section \S\ref{sec:parsep}, we estimate that less than 20\% of interparticle contacts are close enough for direct contact to be possible.} \NOTEK{A movie showing the thermal rearrangements in the center of the heap in included as Supplemental Data.} 

\NOTEK{The nondimensional number typically used to quantify the transition from thermal to athermal behavior is the  P\'eclet number $\mathrm{Pe}$, calculated as the ratio of the particle self-diffusion time $\tau_D=d^2/D$ to the particle advection time $\tau_a=d/v$. 
In previous work \cite{Ortiz2013}, we found that the P\'eclet number at the heap interface, $\mathrm{Pe}_\infty = \frac{d v_\infty}{D_\infty}$, is a good predictor for the formation of a microsphere heap from a dilute suspension. 
 For the 530 nm spheres, we find $D_\infty$ to be 0.55~$\mum^2/s$ from the slope of the long-time mean-squared displacement of microspheres in the absence of external driving, and we measure $v_\infty$ from the streamline bounding the heap.  We calculate that the heaps in this paper probe the range  $\mathrm{Pe}_\infty$ from 4 to 28. In fact, these above-unity values represent an upper bound for the local P\'eclet number. Within the heap, $\mathrm{Pe}$ would be better estimated by the advection time due to the local shear rate $\dot{\gamma}(\phi,y_\perp)$, which depends on both the distance $y_\perp$ into the heap and the packing fraction $\phi$. Therefore, the local P\'eclet number $\mathrm{Pe}(\phi, y_\perp) = \frac{d^2 \dot{\gamma}(\phi,y_\perp)}{D(\phi, y_\perp)}$. Due to hydrodynamic screening, it is likely that an exponential decay in the local shear rate \cite{Brinkman1949, Beavers1967, Ochoa-Tapia1995} causes a similar decay in the local $\mathrm{Pe}$. This effect would dominate 
the presumed algebraic decay of the self-diffusion constant (due to the growth of the shear viscosity at large packing fractions\cite{Krieger1972, 
Brady1993}). Therefore, local values of $\mathrm{Pe}$ are smaller than estimates from $\mathrm{Pe}_\infty$, likely unity or below. Future studies that measure the self-diffusion constant in the interior of the heap would allow for direct probes of the thermal-athermal transition. }


Therefore, we consider the system only in a thermal context, starting from the van der Waals
equation of state $p(A-A_c)= Nk_BT$~\citep{vanderWaals1873}, 
where $p$ is the pressure and $A_c$ is the smallest area that the system can achieve without particle
overlap and $A$ is the area occupied by the heap (including interstitial volume). 
This equation of state is limited to situations where the shape of the accessible volume is
invariant under deformations, and cage-breaking~\cite{Weeks2002} is unlikely. 
Because the degrees of freedom attributed to the particles within an aqueous
suspension do not contribute significantly to the heat capacity, we also assume
that our experiment is performed isothermally. Note that the applicability of the van der Waals 
equation of state to driven granular systems has previously been observed in both
simulations \cite{Luding2001} and experiments \cite{Nichol2012}.

We consider the 2D isothermal deformation of a pack of disks subject to the van der Waals equation of state:
\begin{equation}\label{eqn:vdWeos}
 \sigma\left(A-A_c\right)=C=\text{constant}
\end{equation}
where $\sigma$ is the stress, $A$ is the area occupied by the material, $A_c$ is the excluded area due to the size of the particles such that $A\ge A_c$. A total differential of Eq.~\ref{eqn:vdWeos} gives:
\begin{equation}
 \frac{d\sigma}{dA}=-\frac{\sigma}{A-A_c}
\end{equation}
The bulk modulus $K=-A\frac{d\sigma}{dA}$ is
\begin{equation}
 K=\frac{\sigma}{1-\frac{A_c}{A}}
\label{eqn:K_HS}
\end{equation}
In a material composed of monodisperse particles, the packing fraction $\phi\equiv \frac{NA_1}{A}$, where $N$ is the number of disks, $A_1$ is the cross-sectional area of a single particle. Then, $\frac{A_c}{A}=\frac{\phi}{\phi_c}$, where $\phi_c$ is the densest possible packing fraction. 
Based on our observations in Fig.~3, we define $\alpha$ by Eq.~2 of the main text as
\begin{equation}\label{eqn:alphadef}
 K=\alpha\sigma_0\,\,.
\end{equation}
Combining Eqns.~\ref{eqn:K_HS} and \ref{eqn:alphadef} allows the identification
\begin{equation}\label{eqn:alphadef2}
 \alpha=\frac{1}{1-\frac{A_c}{A}}\,\,.
\end{equation}

To find the stress-strain relation, we consider two instances of Eq.~\ref{eqn:vdWeos}, at steady state ($\sigma_0$, $A_0$) and after deformation ($\sigma_0+\Delta\sigma$, $A_0+\Delta A$):
\begin{align}
  \sigma_0\left(A_0-A_c\right)&=C\label{eqn:vdWsteady}\\
  \left(\sigma_0+\Delta\sigma\right)\left(A_0+\Delta A - A_c\right)&=C\label{eqn:vdWdeform} 
\end{align}
Taking the ratio of Eq.~\ref{eqn:vdWdeform} and Eq.~\ref{eqn:vdWsteady} eliminates the unknown constant $C$. Solving for $\frac{\Delta A}{A_0}$ gives
\begin{equation}
 \frac{\Delta A}{A_0}=\left(1-\frac{A_c}{A_0}\right)\left(\frac{1}{1+\frac{\Delta\sigma}{\sigma_0}}-1\right)
 \label{eqn:gamma}
\end{equation}
Identifying the factor $\left(1-\frac{A_c}{A_0}\right)$ as $1/\alpha$ simplifies the expression to:
\begin{equation}\label{eqn:datafit}
\frac{\Delta A}{A_0}=\frac{1}{\alpha}\left(\frac{1}{1+\frac{\Delta\sigma}{\sigma_0}}-1\right),
\end{equation}
which we use for making fits to the data.
We denote $\alpha$ as the fit parameter to Eq.~\ref{eqn:alphadef}, and $\tilde \alpha$ as the fit parameter to Eq.~\ref{eqn:datafit}.

To test the agreement between this function and Fig.~\ref{fig:lindensity}a, we individually fit each of the seven stress-strain curves to Eq.~\ref{eqn:datafit} and take the compressibility factor $\tilde \alpha$ as a free parameter.
We note that $\alpha$ and $\tilde \alpha$ could possibly differ as one is a constant arising from the geometry, and the other is a fitting parameter to experimental data.
The resulting seven curves are shown in Fig.~\ref{fig:lindensity}c, with the fit values of 
$\tilde \alpha$ provided in the inset. We find
that Eq.~\ref{eqn:datafit} is in good agreement with all compressive data and
small decompressions, while large decompressions ($\Delta\sigma / \sigma_0
\lesssim -0.7$) deviate significantly. Furthermore, the values of $\tilde
\alpha$ obtained in the fits are in agreement with the one obtained in the
linearized data (Fig.~\ref{fig:modulus}c)\NOTEC{, indicating we have achieved a valid measurement of the compressibility factor.  
The magnitude of this compressibility factor indicates significant deviation from 
ideal gas behavior, where  $\alpha \sim {\cal O}(1)$. Expressing the compressibility factor as a virial cluster expansion of packing fraction $\phi$ gives $\alpha = 1 + \sum_{i=1}^\infty B_{i+1} \phi^i$.  Comparing our measured $\alpha$ to the known virial expansion coefficients~\cite{Kolafa2006} indicates the expansion must include terms of at least fifth order. This observation implies that the elastic process measured by the compressibility is subject to significant contributions from many-body interactions.}   In addition, we observe that $\tilde
\alpha$ decreases slightly with increasing $\sigma_0$. This may be associated with 
a stress-dependent packing fraction, an idea which is supported by Fig.~\ref{fig:g}cd \NOTER{and Appendix~\ref{sec:hyd}.}

\section{Conclusions}

The agreement between our observations and the stress-strain relation derived from a simple thermal (van der Waals) equation of state is an encouraging first-step for building more general physical models in this important regime where the P\'eclet number is close to unity. 
This finding highlights the importance of geometric exclusion in setting the material properties of particulate matter.
\NOTEC{We have shown} \NOTER{that} \NOTEC{microsphere heaps support external stress by measuring their bulk modulus.} \NOTER{We have further} \NOTEC{ shown that this stress is supported by the particle network via a mechanism involving many-body particle interactions. }
A direct measurement of the particle positions during deformation 
would allow for a more detailed understanding of the microscopic origins of the stress-strain relation.
The reversibility of the elastic response opens new possibilities for the manipulation of cells and droplets in microfluidic devices, including the design and assembly of new material.




\section{Acknowledgements} We are grateful for support from the National
Science Foundation through an NSF Graduate Fellowship, and grants DMR-0644743
and DMS-0968258. This work was performed in part at the Cornell NanoScale Facility, a member of the National Nanotechnology Infrastructure Network, which is supported by the National Science Foundation (Grant ECCS-0335765).
This work was also performed in part at North Carolina State University facilities: Nanofabrication Facility, Advanced Instrumentation Facility, and Education and Research Laboratory.

\appendix

\begin{figure*}[t]
\centerline{\includegraphics[width=0.75\linewidth]{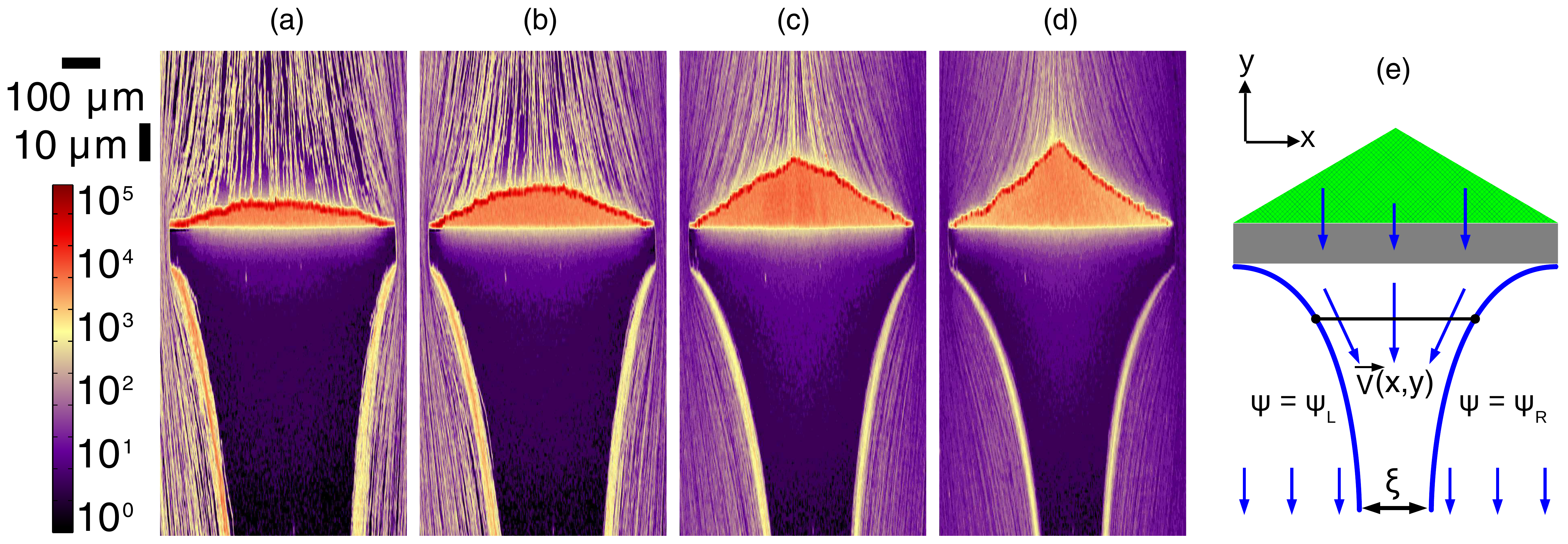}}
\caption{ (a-d) Processed images showing amplitude of intensity fluctuations at different inflow velocities. Each pixel is given an RGB color proportional to the standard deviation measured over 8 s (80 images) after steady state was reached. The colormap is reported in arbitrary units, with purple indicating low fluctuations and red indicating high fluctuations. The large purple areas are the excluded zones. Tracks of single particles are yellow trails. The strongly fluctuating (fluid-like) areas at the surface of each heap are deep red, while the weakly fluctuating (solid-like) interior of the heap is light red. From left to right, the inflow velocities are 3.6~${\mu}\text{m}$/s, 7.1~${\mu}\text{m}$/s, 14.2~${\mu}\text{m}$/s, 35.5~${\mu}\text{m}$/s. Microsphere areal density far from the heap is constant $\rho_\infty=110/(100~{\mu}\text{m})^2$.  (e) Schematic drawing showing streamlines $\psi_L$ and $\psi_R$ bounding the excluded zone of width $\xi$. }
\label{fig:xidefine}
\end{figure*}

\section{Flow Model}\label{sec:hyd}

This section describes the flow model we use to determine the stress $\sigma_0$ supported by the microsphere heap, as a function of $P_0$, the pressure applied at the microchannel inlet. The model is motivated by the observation that \NOTER{there exists} an excluded zone downstream of the barrier, into which particles do not flow (see Fig.~\ref{fig:xidefine}).  This excluded zone
is formed by liquid which has flowed through the heap and barrier. Given knowledge of the channel dimensions, we find that the asymptotic width $(\xi)$ of this zone measured far downstream of the barrier, together with the flow velocity measured far upstream of the barrier, provides a measure of both the flow rates around and through the heap/barrier, and the applied stress on the heap.

In making this calibration, we assume that the flow is laminar, and is plug-like along the $x$-direction far away from the heap. Following \cite{Acheson1990}, we calculate  the thin-film Reynolds number from the channel geometry and maximum fluid speed ($U=100~{\mum}$/s) as
$Re=\frac{\rho U L}{\eta}\left(\frac{H}{L}\right)^2 \approx {\cal O}(10^{-8})$, 
with $\rho$ the density and $\eta$ the viscosity, validating the assumption of laminar flow. 
We model the fluid flow within the microchannel using Stokes'
equation $\eta\vec\del^2\vec{u}=\vec\del p$, where $p$ is the pressure, and the the incompressibility condition
$\vec\del\cdot\vec{u}=0$. We numerically verify that the flow within the microchannel without heap and barrier deviates by less than 10\% from a plug-like flow in the $(x,y)$-plane except within two particle diameters from the side walls (similar to \cite{Darnton2001}). We neglect the $z$-direction, where we expect Poiseuille flow in microchannel areas without a heap, and plug-like flow within the heap due to Darcy's law.

\begin{figure}
\centering
\includegraphics[width=\columnwidth]{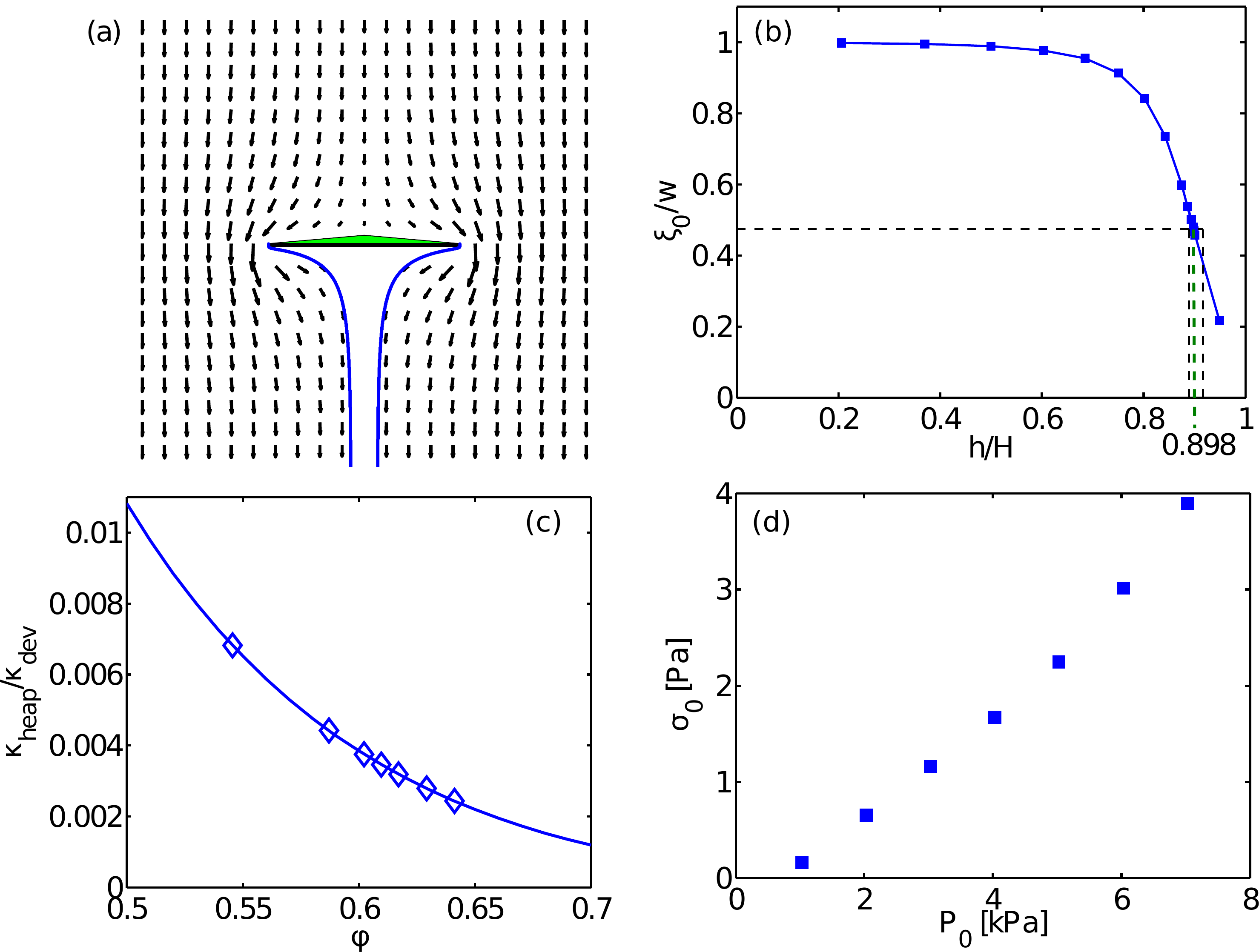}
\caption{(a) Simulated depth-averaged velocity field. The two solid lines are the streamlines that define $\xi$. (b) Solid line shows plot of $\xi_0/w$  in simulation versus the value of $h/H$.  Dashed lines show the range of experimental measurements of $h/H$ and $\xi_0/w$.  (c) Simulated relative heap permeability versus heap packing fraction $\phi$ at 7 values of $P_0$ (left is low $P_0$, right is high). Solid line given by Eq.~\ref{eqn:CK}. 
(d) Final calibration of $\sigma_0$ (pressure on heap surface calculated by Eq.~\ref{eqn:sigmaAve}) as a function imposed pressure $P_0$ for each of the 7 experimental conditions.}
\label{fig:impedanceVsimul}
\end{figure}

The flow everywhere in the $xy$-plane can therefore be modeled in a fashion similar to Darcy flow since the viscous stresses in the $z$-directions are far larger than those in the $x$ and $y$ directions within the microchannel. We solve the two-dimensional Laplace equation for the pressure, $\vec\del\cdot\left[ \kappa(x,y) \vec\del p \right]=0$ using the MATLAB PDEtool. The local permeability $\kappa$ is chosen such that the fluid volume flux $\vec{v}(x,y)=-\frac\kappa\eta\vec\del p(x,y)$ agrees with experimental streamline observations.  $\vec v(x,y)$ models the mean liquid velocity in the microchannel regions. 
We solve the two-dimensional Laplace's equation for the pressure . Each of three regions of space (barrier, heap, free channel) is assigned a constant value of $\kappa$ ($\kappa_\text{bar}$, $\kappa_\text{heap}$, $\kappa_\text{dev}$, respectively). A sample such solution is shown in Fig.~\ref{fig:impedanceVsimul}a. 

Using the pressure distribution $p(x,y)$ within the channel, we obtain streamlines of constant $\psi(x,y)$ satisfying $\vec v(x,y)=\left(\frac{\partial \psi(x,y)}{\partial y},-\frac{\partial
\psi(x,y)}{\partial x}\right)$. We are particularly interested in the streamlines touching the left and right edges of the barrier, labeled $\psi_L$ and $\psi_R$ in Fig.~\ref{fig:xidefine}e, respectively, as these two streamlines form the boundaries of the excluded zone. Particles from streamlines that intersect the barrier are either added to the heap or travel along the edge of the surface until they flow off the surface of the heap at the ends of the barrier. Thus, they follow the pair of streamlines from the ends of the barrier, and leave a particle-free (excluded)
zone at the center of the flow. Note that, at finite Pe, particle diffusion does allow some particles to cross streamlines and enter this zone.

Downstream of the barrier, $\psi_L$ and $\psi_R$ asymptote to a lateral separation distance $\xi$ as the plug-like flow is restored where streamlines are parallel, and the volume flux $\vec v$ is independent of $x$.  We find the ratio of flow rates through the 
heap/barrier ($Q_\text{heap}$)
and the total flow volume through the device ($Q_\infty$) via a ratio of the widths of the excluded zone and the entire device: 
$\frac{Q_\text{heap}}{Q_\infty}=\frac{\xi}{W}$. In the following, we match simulated shapes of $\Psi_L$ and $\Psi_R$ to the experimentally obtained shape of the excluded zone. To determine the average flow velocity upstream of the barrier, we measure the peak of the spatial cross-correlation of particle trajectories in adjacent frames.  $Q_\infty$ is then  detemined from the particle velocity far upstream from the heap, as is the pressure gradient far away from the heap.

We first compare simulation and experiment for a device in the absence of a heap on the barrier. 
The analytic expression for the relative permeability 
of the barrier is 
$\frac{\kappa_\text{bar}}{\kappa_\text{dev}} = \left(1-\frac{h}{H}\right)^2$,
where $h$ is the height of the barrier within the channel of height $H$  (both in the $z$-direction). Fig.~\ref{fig:impedanceVsimul}b shows the simulated $\xi_0/w$ as a function of $h/H$. We measured $h=0.84~\mu$m and $H=0.94~\mu$m for our device. The flow shapes in experiment and theory closely resemble each other, and we find that our measured $\xi_0/w$ agrees with the simulated $\xi_0/w$ for our measured $h/H$ in Fig.~\ref{fig:impedanceVsimul}b.

For devices with heaps, we adjusted $\kappa_\text{heap}$ to yield a simulated flow profile equal to that observed in experiments \NOTER{while maintaining the ratio of $\kappa_\text{bar}/\kappa_\text{dev}$ obtained from Fig.~\ref{fig:impedanceVsimul}b and a shape of the heap that is observed in experiments}.  We find that $\kappa_\text{heap}$ is lower for heaps assembled at higher steady-state stresses, and higher for assembly at low stresses. However, we had to assume a uniform $\kappa_\text{heap}$ for the heap at each flow condition due to the insufficient information on stream lines within the heap. In view of the position-independent strain described in the previous section, this choice appears not to be problematic.

We now compare the permeabilities obtained from the combination of experiment and numerical model to those predicted by the Carman-Kozeny model for a collection of
spheres of packing fraction $\phi\equiv \frac{V_\text{spheres}}{V_\text{total}}$ (\cite{Ergun1949,Carman1937}). Given $\kappa_\text{dev}\approx\frac{1}{12}H^2$,
the Carman-Kozeny relation relates the permeability of the heap to its packing fraction $\phi$:
\begin{equation}\label{eqn:CK}
\frac{\kappa_\text{heap}}{\kappa_\text{dev}}=\frac{(1-\phi)^3}{\phi^2}\frac{d^2}{180}\frac{12}{H^2}
\end{equation}
The solid line in Fig.~\ref{fig:impedanceVsimul}c corresponds to the Carman-Kozeny
permeability at a given apparent packing fraction. \NOTEC{Note, these apparent packing fractions are not confirmed by direct measurements, only inferred from the model.} The apparent packing fractions are in the range 0.54 to 0.64, although we note that the  nearly 2-dimensional packing renders the void volume of a different shape that the 3-dimensional Carman-Kozeny system. 
Importantly,  Fig.~\ref{fig:impedanceVsimul}c highlights the need for detailed structural information to better understand the nature of the rigidity, because the stress supported by the heap depends on this coupling between its packing faction and its permeability. 

Using our experimentally-determined permeabilities, we use the full solution of the Laplace equation to determine the fluid stress on the heap, without needing specify the local packing fractions. Inspection of the data set shows that the flow velocity within heap and barrier is approximately in the $y$-direction.  \NOTER{Together with the incompressibility of the liquid, this implies that the} flow velocity through barrier is well-defined as a function of only $x$. Since the observable chosen for the deformation is the area, we identify the fluid stress as the area-averaged stress formed by 
\begin{equation}\label{eqn:sigmaAve}
\sigma=\frac 1 A \int^{w/2}_{-w/2}\left(p\left(x,y_\text{surf}(x)\right)-p\left(x,y_\text{bott}\right)\right)\left(y_\text{surf}(x)-y_\text{bott}\right)\text{d}x\,\,.
\end{equation}
Here $y_\text{surf}(x)$ is the location of the upstream surface of the heap, and $y_\text{bott}$ is the location of the interface between heap and barrier. The resulting calibration curve for all data in this publication is shown in Fig.~\ref{fig:impedanceVsimul}d.


\end{document}